# Performance Analysis of Software to Hardware Task Migration in Codesign

**Dorsaf SEBAI[1,2], Abderrazak JEMAI[1,2] and Imed BENNOUR[3]**

**[1] LIP2 Laboratory, Faculty of Science of Tunis, Tunisia**

**[2] INSAT, B.P. 676, 1080 Tunis Cedex, Tunisia**

**[3] EμE laboratory, Faculty of sciences at Monastir, Tunisia**

**Abstract**

The complexity of multimedia applications in terms of intensity of computation and heterogeneity of treated data led the designers to embark them on multiprocessor systems on chip. The complexity of these systems on one hand and the expectations of the consumers on the other hand complicate the designers job to conceive and supply strong and successful systems in the shortest deadlines. They have to explore the different solutions of the design space and estimate their performances in order to deduce the solution that respects their design constraints. In this context, we propose the modeling of one of the design space possible solutions: the software to hardware task migration. This modeling exploits the synchronous dataflow graphs to take into account the different migration impacts and estimate their performances in terms of throughput.

## 1. Introduction

The enhancement of multimedia applications reaches its culminating point because of the growing consumers needs in all domestic and professional audio video domains. To answer these needs more and more rigid, the embedded systems rapidly evolve towards multiprocessor systems on chip (MPSoCs) particularly those based on networks on chip (NoCs) as communication architecture. The number of processors per chip, the diversity of their types as well as their communications complicate the MPSoCs design; without forgetting the multimedia applications complexity in terms of computation intensity and data abundance and heterogeneity. So, the principal challenge of designers is to face this NoC-based MPSoC design complexity and provide robust systems in the shortest delays. To deal with these conflicting design challenges, designers have to estimate principal characteristics of the final system early in the design process of MPSoCs; which results in a final implementation where productivity and quality are simultaneously guaranteed.

Designers must control the ever growing MPSoC Design Space Exploration (DSE) where different choices are investigated in order to determine the appropriate choice that leads to a fair compromise between the different conflicting design objectives. Typically, the performance estimation is an important part of the DSE. Different choices of the application parallelization, the target platform and the mapping of the application onto the platform need to be estimated in terms of different quality criteria. If the constraints (energy consumption, throughput, etc.) drawn by the designers are not achieved, modifications should be brought to the application decomposition and/or the platform and/or the proposed mapping in order to find an MPSoC configuration that meets the designers constraints.

In recognition of the growing need to the MPSoC performance estimation, different approaches aim at estimating the overall system performance. In [1], three approaches are defined. First, the simulation-based approach is based on an evaluation of the system behavior by means of simulation (native execution, Instruction Set Simulator, etc.) in different abstraction levels. Kai Huang and al [2] exploits the Simulink platform to simulate the multimedia applications on different hardware platforms. The H.264 decoder is used as a case study to validate this work. Its simulation on different platforms (change of the processors type and number) estimates the number of consumed cycles per processor for execution and communication. In the same way, Teresa Medina Leon [3] proposes the MJPEG decoder simulation on the MiniNoC platform to estimate the time required for a frame decoding. The MiniNoC platform, implemented in C++, simulates in the register transfer level a platform composed of four mini MIPS processors displayed in four nodes that communicate with each other via a mini NoC composed of four routers. Second, the trace based approach consists in collecting the application execution traces. Designers operate a single simulation at the beginning of the design phase from which they extract traces of the application execution on the target platform. These traces can be, for instance, the size of the transferred data on the communication platform, the number of transactions between every pair of tasks, the execution times of different tasks, etc. The collected traces are organized in the form of a Communication Analysis Graph (CAG). The CAG analysis allows the designers to produce several statistics about the system performances. The trace based approach is generally used when the initial simulation is difficult to reproduce. This approach was exploited in [4] to lead to the optimal communication mapping on a predefined target architecture assuming that the application is already partitioned and mapped. Traces collected in this work serve at calculating, for every edge of the CAG, a weight that reflects the frequency, the volume and the criticism of transactions between every communicating unity. Finally, the static approach that tries to avoid the computationally prohibitive and exhaustive simulation, makes use of "static" models such as graphs, mathematical equations, UML components [5] and XML tags to estimate the MPSoCs performances.

In this paper, we opt for static approach using exactly the Synchronous DataFlow Graphs (SDFGs) to model applications as well as their mapping on target platforms. SDFGs are extremely used for MPSoCs performance estimation since they fit well with the characteristics of streaming multimedia applications. Moreover, they can model many mapping decisions of an application on NoC-based MPSoC adding new actors and edges to the initial SDFG of the application. Among the several design flows [6] [7] [8] that make use of SDFGs as a model of computation, we focus in this paper on the predictable design flow established by Sander Stuijk [9]. These design flows do not model the migration of the application software tasks to hardware ones. They just assume that executing tasks on hardware blocks requires half the number of cycles as executing them on general purpose processors.

So, the principal motivation that governs this work is the consideration of one of these DSE alternatives: the software to hardware task migration. The migration performance estimation refines the choice of the optimal solution from the design space. It is a solution adopted to face part of the design problems; using a hardware block instead of a software task running on a general-purpose microprocessor which is not fast enough to achieve design goals. If the migration solution of a given task respects the constraints fixed by the designers, this solution will be adopted, otherwise the migration of other tasks can be performed to satisfy the design constraints. In this context, we begin by defining the impacts that are caused by the migration of a software task to a hardware one. Second, we opt to model these impacts using SDFG. The latter takes charge of application modeling in the form of graphs from which designers can estimate the application throughput.

The paper is organized as follows. The next section introduces the MPSoCs performance estimation using SDFGs. Section 3 details the software to hardware task migration impacts as well as their SDFG modeling. The case study of the MJPEG decoder is given in Section 4. Conclusions and perspectives are drawn in Section 5.

## 2. MPSoC performance estimation by SDFGs

Seeing that this paper focuses on the MPSoCs performance estimation using SDFGs, we will first introduce the MPSoCs as well as their architecture and provide an overview of the most important SDFGs properties. Then, the predictable design flow will be presented as a flow case that uses SDFGs to estimate the performance of multimedia applications, mapped on MPSoCs, in terms of throughput.

### 2.1 Multiprocessor systems on chip architecture

The increasing and exponential complexity of multimedia applications has widely promoted the use of SoCs composed of several processors. The processors number and diversity per chip require a powerful communication infrastructure that supports numerous transactions between processors. Therefore, bus based SoCs have rapidly induced bottlenecks and led to the NoC emergence for their flexibility and scalability. The NoC based MPSoC architecture is composed of a set of tiles connected via the NoC. Every tile is formed by a processor (P), a local memory (M), a network interface (NI) that connects tiles to the NoC and a communication assist (CA). The latter is responsible for the data transfer between the local memory of the tile and the network interface.

The NoC is a set of routers connected to each other by links according to a determined topology. Routers are referenced by two addresses X and Y reflecting their respective positions in two dimensions width and length. In addition, every router is connected to its immediate neighbors in the two dimensions X and Y.

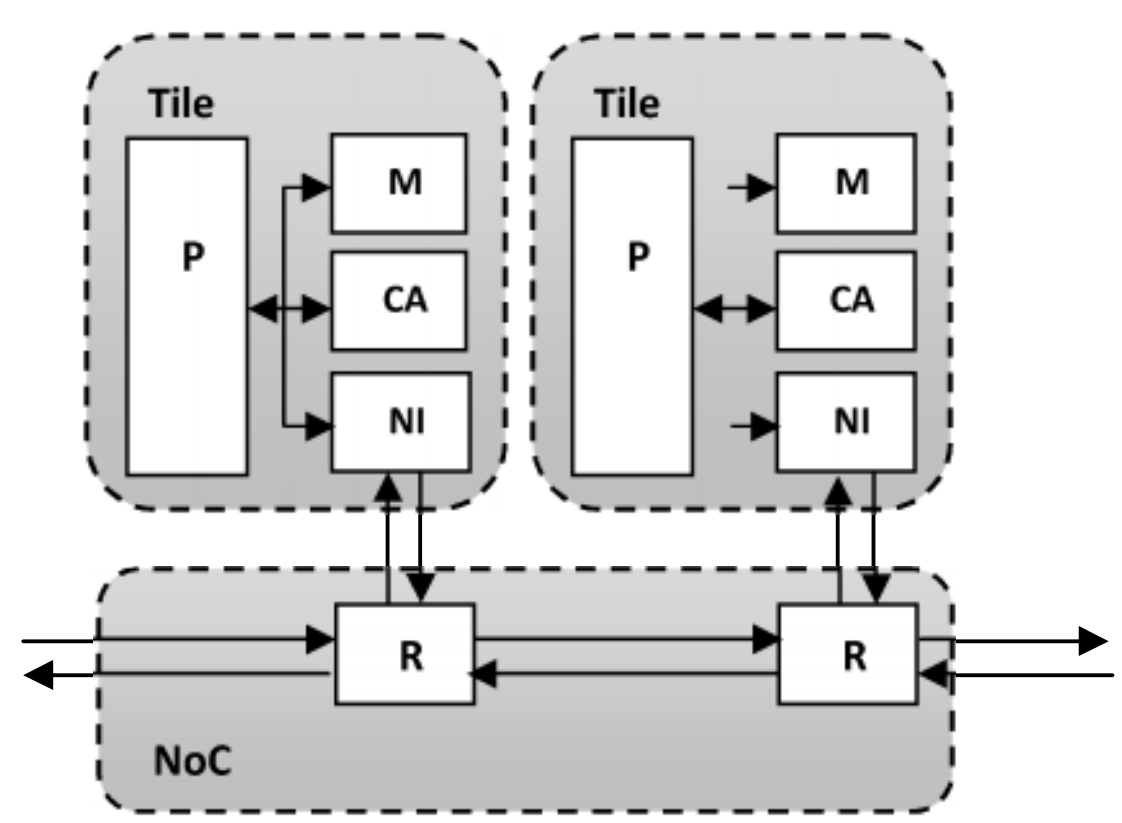

Fig. 1 NoC based MPSoC architecture

## 2.2 Synchronous dataflow graphs

SDFG, proposed by Lee and Messerschmitt [10], is a model of computation used to model multiprocessor applications and analyze their temporal behavior during design phase. Fig. 2 illustrates a simple SDFG example formed by two actors A and B. Each actor presents an application task having a given execution time. Edges between A and B model data communicated between each other. We have to mention that the self-edge of actor B as well its initial token allow the control of the actor internal state. It assures that actor B can only begin its next execution when it finishes its current one. In SDFGs, data are communicated in the form of tokens which is a data container where a fixed amount of data can be saved. For instance, edge from B to A contains 2 initial tokens used to launch the SDFG execution.

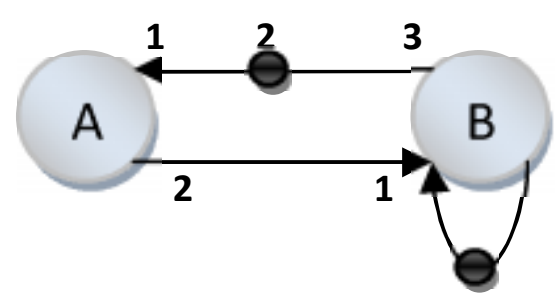

Fig. 2 SDFG example

Numbers in the extremities of the arcs, named rates, designate the number of tokens consumed and produced by an actor during its firing. In the SDFG example of Fig. 2, actor A consumes 1 token and produces 2 tokens whereas actor B consumes 1 token and produces 3 tokens. Since default rates are equal to 1, rates “1” can be not mentioned in the SDFG to avoid the graph obstruction.

A fundamental property of SDFGs is that every time an actor fires it consumes the same amount of tokens from its input ports and produces the same amount of tokens on its output ports. In addition, an actor can only begin its execution when tokens necessary for its firing are available in all its input edges. Actor A of the SDFG example fires as soon as at least 1 token is available in its incoming edge (edge from B to A).

SDFGs are widely exploited to derive the applications throughput using analytical methods. SDFG throughput is formally defined as the average number of the SDFG iterations per time unit. Literature evokes two equivalent methods of SDFG throughput computation: Maximal Cycle Mean [11] and self-timed execution [9] methods.

## 2.3 Case of predictable design flow

The predictable design flow [9] enables designers to map an application SDFG on NoC based MPSoC platform while respecting a throughput constraint and minimizing the platform resources usage in terms of processors, memories and bandwidths. The flow takes as entrance the application SDFG, the throughput constraint fixed by the designer and the target platform. It generates an MPSoC configuration which proposes the mapping of the application SDFG actors on the target platform as well as the scheduling of their communications. The open source tool SDF3 [12] is the implementation of the predictable design flow. It also implements algorithms for the SDFGs generation, visualization, transformation and analysis.

The predictable design flow consists of four phases. The « memory dimensioning » phase takes an interest in memory allocations of all edges in the application SDFG. The second phase « constraint refinement » specifies the constraints that edges must respect in terms of latency and bandwidth. « Tile binding and scheduling » phase proposes the mapping and scheduling of all actors in the application SDFG on the target platform tiles while respecting the throughput constraint. The last phase « NoC routing and scheduling » takes charge of communications mapping and scheduling on the platform NoC. Each phase produces a SDFG that models phase decisions adding new actors and edges to the initial application SDFG. We detail the memory-aware and the binding-aware SDFGs generated respectively by the first and third phase.

The memory-aware SDFG models the memory allocations made by the first phase of the flow. Actors of which tokens size exceeds the storage capacity of tiles local memories will be stored in shared memory tiles. During its firing, actor B of Fig. 3a consumes *i* tokens and produces *o* tokens. Consumed tokens have a large size and must be stored in a remote memory. So, actor B must operate remote accesses to bring these tokens in case of need. These remote accesses to the remote memory are modeled as shown in the SDFG of Fig. 3b.

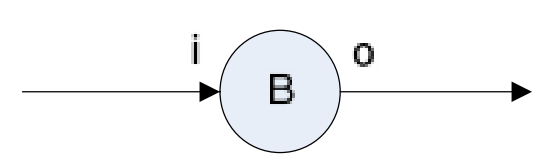

(a) Actor B before memory allocation

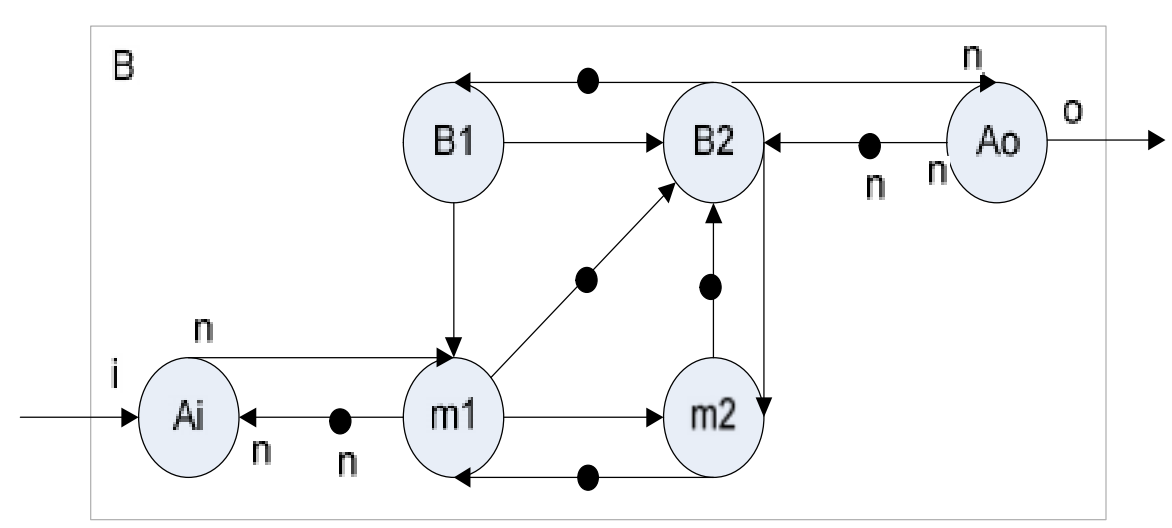

(b) Actor B after memory allocation

Fig. 3 Memory-aware SDFG [9]

Actor B is henceforth modeled by two actors B1 and B2. The remote memory is modeled by two actors m1 and m2. While actor B2 operates its $(i)^{th}$ execution, actors B1 and m1 pre-fetch data necessary for the $(i+1)^{th}$ execution of actor B2 from the remote memory. Actors B2 and m2 are in charge of actor B execution. They also fetch data from the remote memory when the pre-fetch phase cannot look for all data necessary for the $(i+1)^{th}$ execution of actor B2. Actors Ai and Ao as well as their cycles with actors m1 and B2 reserve respectively the input and output behavior of actor B. As for number *n*, it represents the number of times actor B fires. When *n* tokens are available in the entrance of actor Ai, it fires. Its execution produces *n* successive executions of actors B1, B2, m1 and m2. These *n* executions provide *n* tokens in the entrance of actor Ao what induces its firing. The Ao execution allows then next *n* successive executions of actors B1, B2, m1 and m2.

The binding-aware SDFG models mapping decisions of the predictable design flow third phase. Edges whose source and destination actors A and B are mapped to the same tile are modeled with a back-edge from A to B. As shown in Fig. 4a, the back-edge has $\alpha_{tile-n}$ initial tokens which present the remaining free memory space for the communication between A and B. We notice that $\alpha_{tile}$, $\alpha_{src}$ and $\alpha_{dst}$ parameters are computed by the first phase of the flow. $\alpha_{tile}$ designates the memory, in tokens, required for the communication between A and B when they are mapped to the same tile. $\alpha_{src}$ and $\alpha_{dst}$ designate the memory required respectively in source and destination tiles when A and B are mapped to different tiles.

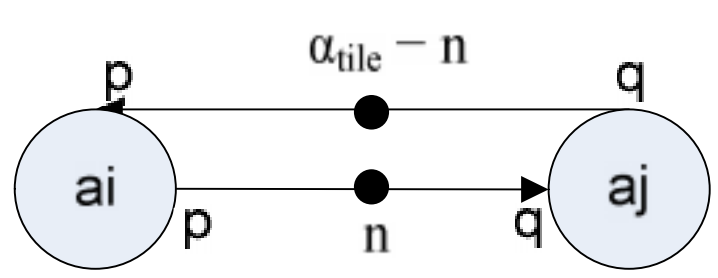

(a) Actors A and B mapped on the same tile

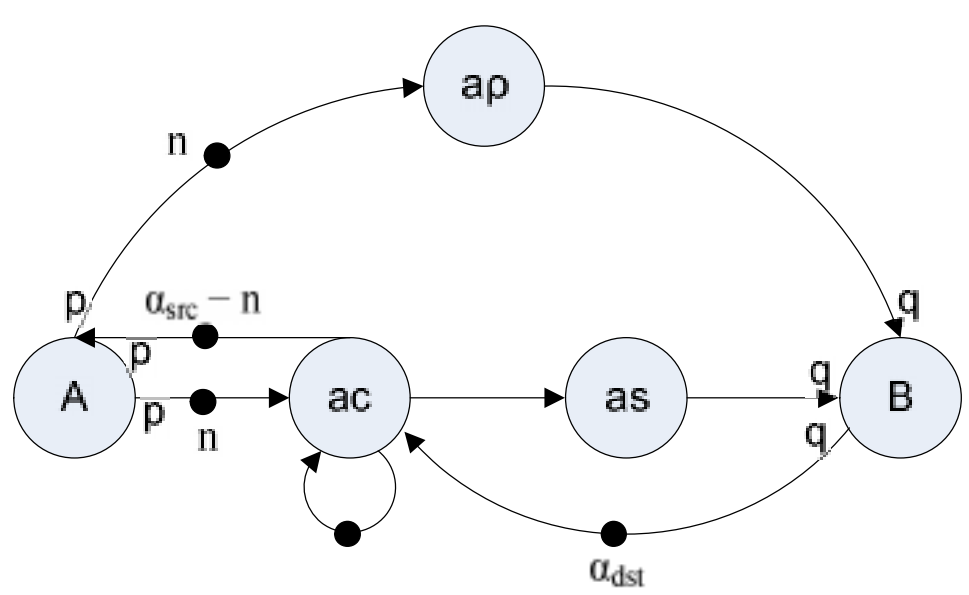

(b) Actors A and B mapped on different tiles

Fig. 4 Binding-aware SDFG [9]

Edges of which the source and destination actors A and B are mapped to different tiles are bound to a connection of the platform NoC as shown in the Fig. 4b. The remaining free memory spaces for the communication between A and B in the source and destination tiles are respectively modeled by the back-edges from $a_c$ to A and from B to $a_c$. Actor $a_c$ models the sending latency of a token via the connection. Actor $a_\rho$ presents the minimal latency, calculated by the second phase of the flow, between the production and the consumption of a token communicated between A and B over the connection. Actor $a_s$ considers the worst case where a token necessary for the execution of actor B arrives at the end of its TDMA time slot. Therefore, actor $a_s$ models the time that actor B has to wait to reach its next TDMA slot and consume its token.

## 3. Software to hardware task migration

### 3.1 Migration impacts

Software to hardware task migration must be considered in the precocious design steps to estimate its gain compared to a pure software application. According to this gain, designers will take the decision for or against the hardware implementation of one or several tasks of the application.

• *Impact 1: The hardware task execution time:*

The first migration impact is the remarkable decrease in the execution time of the hardware task compared to the software one. The hardware implementation of the migrated task will be executed on special purpose hardware and integrated circuits to perform the task function; what causes a significant reduction on its execution time when compared to its execution on a general purpose processor. Executing tasks on hardware blocks may require less than half the number of cycles as executing them on processors.

• *Impact 2: The migrated task workload:*

This impact draws attention to the workload of the software task that disappears once it is executed separately on a hardware block. To clarify the idea, we will consider a simple example of two tasks T1 and T2 executed on the same processor with a TDMA period of 100 time units. Before migration, T1 has 30% of the total TDMA period and T2 takes the remaining 70% for its execution. If we consider the worst case in which a token needed to fire task T1 arrives exactly at the end of its time slice, it has to wait the time slice of T2 (70 time units) before it can fire. After migrating T2 from software to hardware, T1 will be the only task executed on the processor and the whole TDMA wheel size of the processor will be at its disposal. Therefore, T1 has no longer to wait the time slice allocated to T2 which speeds-up its firing and obviously its throughput.

• *Impact 3: The migrated task communication:*

One of the important aspects that must be considered in the software to hardware task migration is the type of communication that will be used to transfer data from software tasks to hardware blocks and vice versa. Since we work at the system level, we will not consider the communication details such as data synchronization, wrappers to integrate the hardware blocks, etc. Different types of Software/Hardware (SH), Hardware/Software (HS) and Hardware/Hardware (HH) communications can be distinguished. In this work, we will focus on three communication types that we have named SH1, HS1 and HH1.

Table 1: SH1, HS1 and HH1 communications

| | Description | Archit |
|---|---|---|
| **SH1** | The software task executed on a general-purpose processor sends its output data to the buffer of the hardware block. | 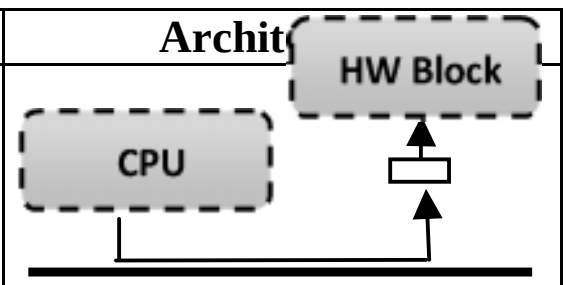 |
| **HS1** | The software task executed on a general-purpose processor reads its input data from the buffer of the hardware block. | 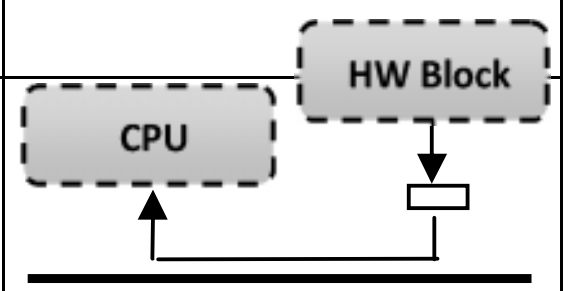 |
| **HH1** | The producing hardware block sends its output data to the buffer of the consuming hardware block. | 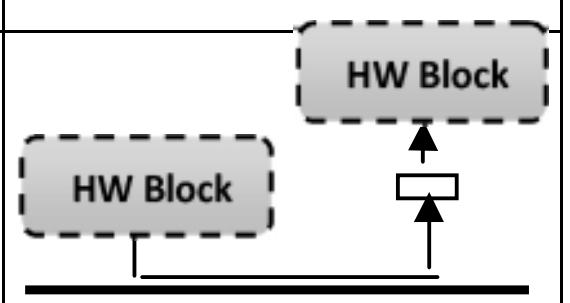 |

• *Impact 4: The communication overhead:*

This impact is the direct result of the hardware task communication with other software and hardware tasks. It depends on the initial mapping of the migrated task before the migration, the mapping of its communicating tasks and the sense of the communication. Therefore, every communication type will be treated apart.

To explicit the case of the SH1 communication overhead, we will first consider the example of two software tasks T1 and T2 mapped on the same tile. Hence, T1 and T2 communicate locally in the memory of the tile and do not need to transfer data via the NoC. If we decide the task T2 migration, tasks T1 and T2 are no more mapped on the same tile and have henceforth to send their communicated data via the NoC. The usage of the NoC leads to an additional load that does not exist before T2 migration.

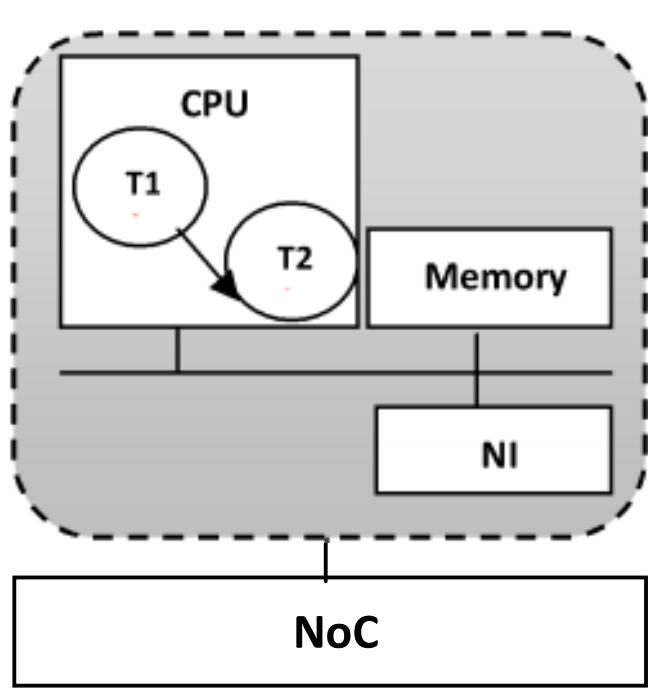

(a) Before migration of T2

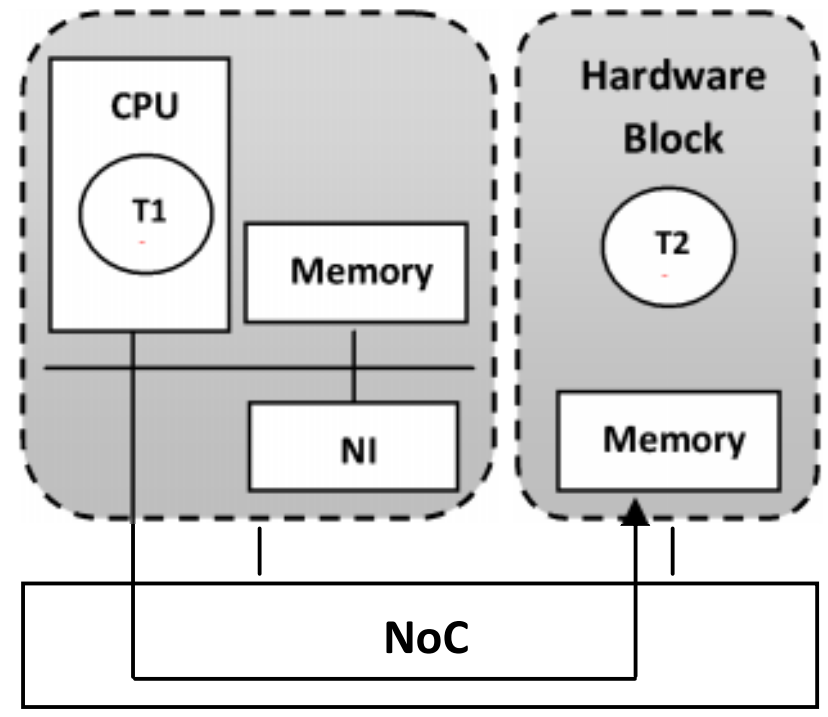

(b) After migration of T2

Fig. 5 SH1 communication with T1 and T2 in the same tile before migration

If we consider that, before migration, tasks T1 and T2 are mapped in different tiles. The migration of task T2 will not cause a communication overhead since tasks T1 and T2 already communicate via the NoC before the migration occurs.

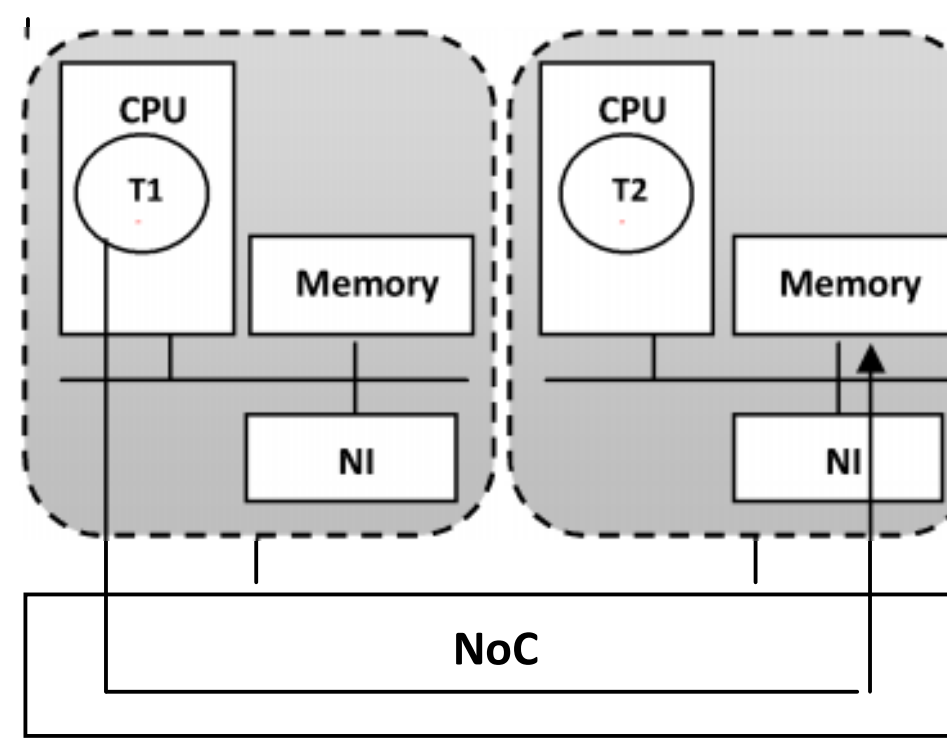

(a) Before migration of T2

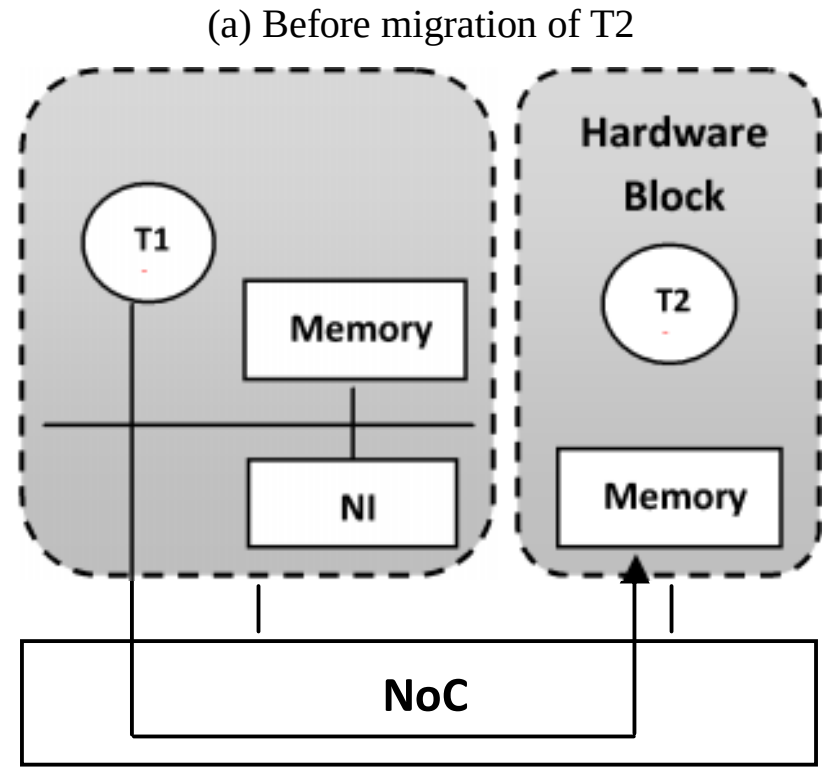

(b) After migration of T2

Fig. 6 SH1 communication with T1 and T2 in different tiles before migration

As already seen in the SH1 communication, the migration of task T2 produces a communication overhead if the initial software tasks T1 and T2 are mapped to the same tile before the migration. If tasks T1 and T2 are bound to different tiles before the migration occurs, a communication overhead takes place. In fact, in the case of the HS1 communication type, the hardware block does not send data to the consuming software task. The hardware task saves its output data in its local memory (1); then the software task must read (2) and bring the data necessary for its firing from the memory of the hardware block (3) as shown in Fig. 7. Therefore, these remote accesses create an extra load of communication via the NoC.

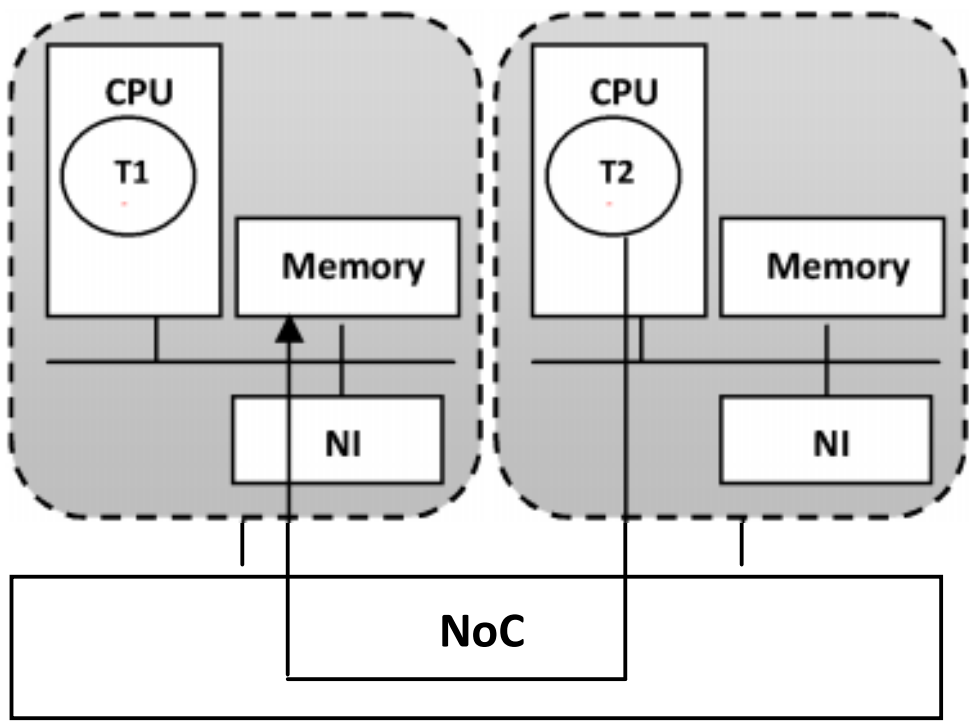

(a) Before migration of T2

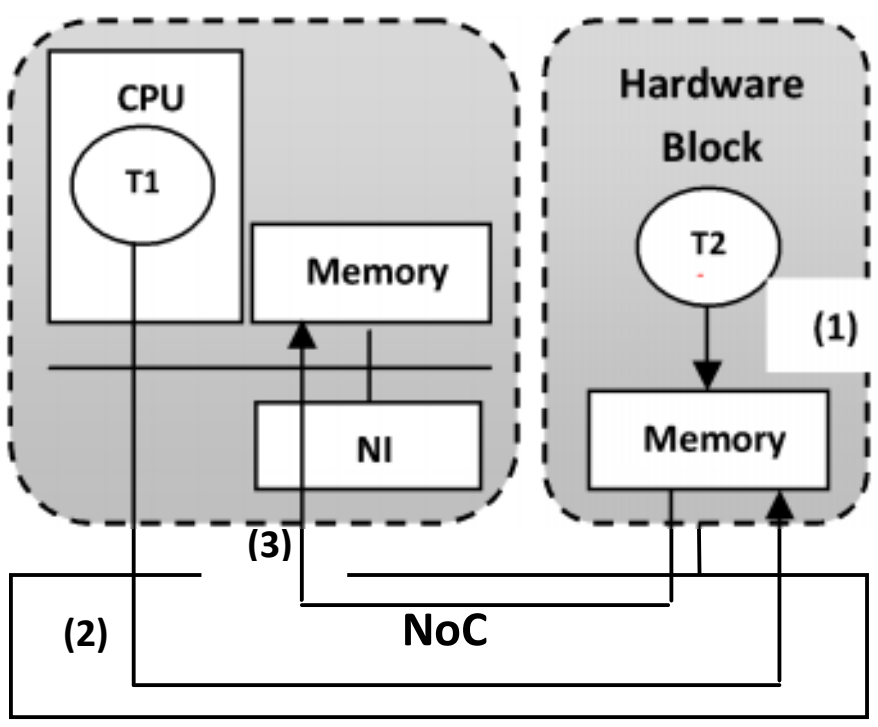

(b) After migration of T2

Fig. 7 HS1 communication with T1 and T2 in different tiles before migration

In the ultimate case, the HH communication overhead, we treat the case of a software task T2 that communicates with a hardware task T1. The migration of task T2 does not engender any extra communication load since T1 and T2 communicate via the NoC before the migration occurs. As shown in Fig. 8, task T2, before migration, sends its data via the NoC to the memory of the hardware block. After the migration takes place, task T2 still sends its output to the hardware block memory.

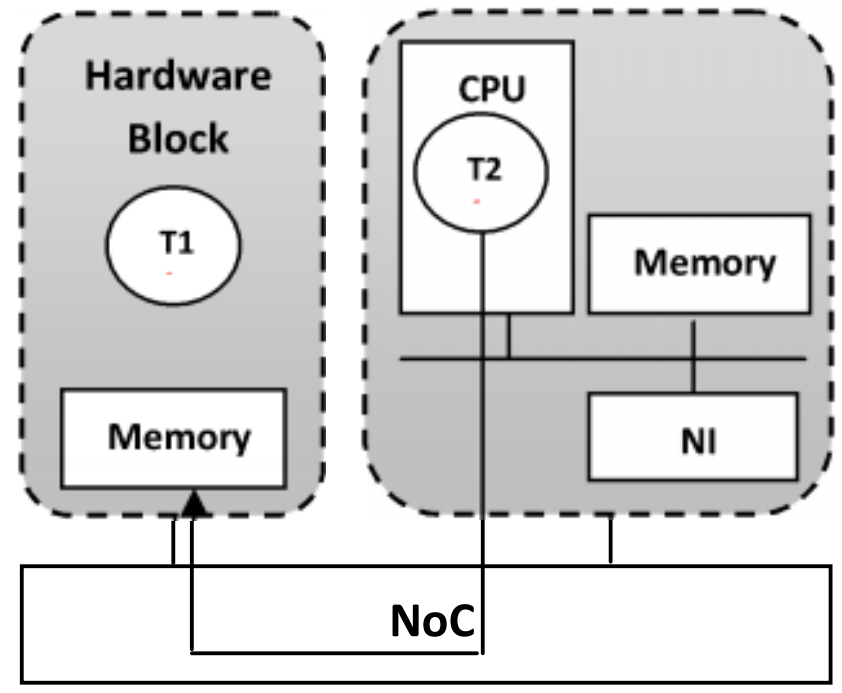

(a) Before migration of T2

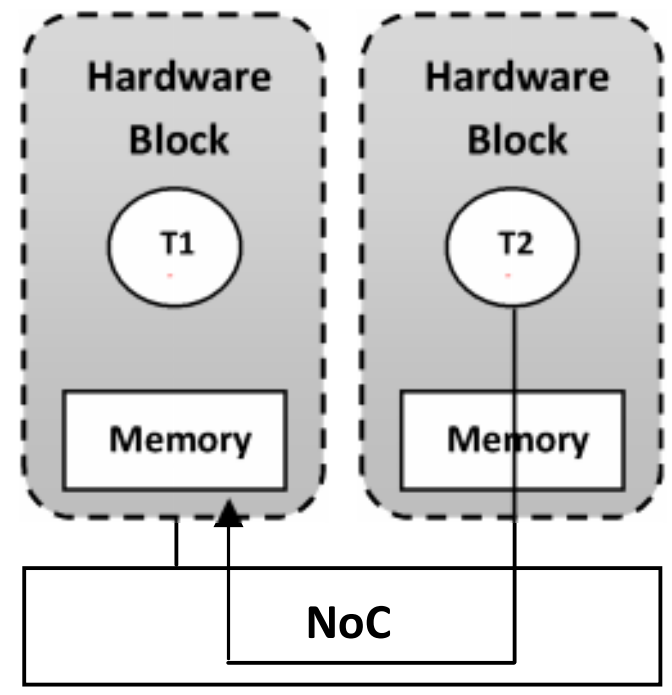

(b) After migration of T2

Fig. 8 HH1 communication

## 3.2 Migration modeling

We will eventually make use of the predictable design flow transformations; so as to have a system-level model that considers all the migration impacts. We have to notice that the communication overhead impact is the direct result of the used communication type. So, the modeling of the migrated task communication impact (impact 3) includes the modeling of the communication overhead impact (impact 4).

• *Impact 1 Modeling:* Concerning the first migration impact, we can act on the execution times of actors that model the hardware tasks in the SDFG. The execution times of actors that model the migrated tasks will be reduced compared to their execution times before migration.

• *Impact 2 Modeling:* In [13], TDMA time slice allocations are modeled by increasing the execution time of every actor firing with the fraction of the TDMA time which is reserved by other actors. It means that the worst case is considered; that is why the firing of a given actor is usually postponed by the TDMA time allocated to other actors. Since the migrated task workload will disappear after migration, we will not consider its TDMA time in the execution time of the software task that was executed on the same processor on which the migrated task was also executed.

• *Impact 3 Modeling:* The binding-aware SDFG, used by the third phase of the predictable design flow to model the SS communication, can be used to model the SH1 and the HH1 communications. In fact, SS, SH1 and HH1 communications use the same principle to transfer data via the NoC:

- In the SS communication, the **software task** T1 sends data to the *local memory* of the tile on which **software task** T2 is mapped.
- In the SH1 communication, the **software task** T1 sends data to the *local memory* of the **hardware task** T2.
- In the HH1 communication, the **hardware task** T1 sends data to the *local memory* of the **hardware task** T2.

As we can notice, in these three types of communication, data are usually transferred from the producing task T1 to the local memory of the consuming task T2. The difference consists in the extremities of the communication either they are software or hardware. Therefore, the SH1 and HH1 communications can be modeled by the same SDFG shown in Fig. 4b.

As explained in Table 1, during an HS1 communication, the hardware task stores its output in its local memory; then the consuming software task operates remote accesses to the memory of the hardware block to fetch data necessary for its firing. This communication type has the same principle of tasks that do not have enough space in their tiles to store their data. So, they have to store their data in a shared memory; then they will bring them in case of need. Hence, to model the HS1 communication, we propose the memory-aware SDFG already illustrated in Fig. 3b.

## 4. Motion JPEG decoder case study

### 4.1 Motion JPEG decoder

The motion JPEG decoder is a multimedia application whose building blocks are used in many image and video processing algorithms. The first block VLD performs variable length decoding. IZZ reorders the stream of pixels coefficients according an inverse zigzag sequence. IQ and IDCT functional blocks respectively operate the inverse quantization and the inverse discrete cosine transform. CC and RE are not specified in the official MJPEG standard [14] but they are necessary to adapt the pixel stream to output peripherals. CC allows color conversion from the YCbCr color scheme to the RGB one. RE reorders the pixels to rebuild the decompressed image.

### 4.2 Mapped MJPEG decoder before migration

Referring to the MJPEG decoder implementation, the resulting SDFG of the MJPEG decoder decoding video frames of resolution 32*24 is shown in Fig. 9. Actors IZZ, IQ, IDCT and CC operate on blocks of 8 by 8 pixels so that a token is equivalent to a block of 8 by 8 pixels. Actors VLD and RE operates on the whole image hence on 12 = (32/8) * (24/8) matrices of 8 by 8 pixels.

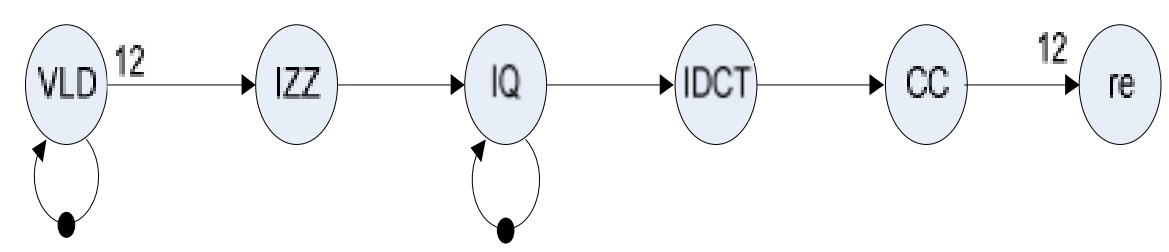

Fig. 9 MJPEG decoder SDFG

The platform on which the MJPEG decoder will be mapped, in this case study, is formed by three tiles T1, T2 and T3 having each one a total TDMA wheel size of 100000 time units. Then, we assume that the VLD and IZZ actors are mapped on the same tile T1, actors IQ and IDCT on the tile T2 and actors CC and RE on the tile T3. The actors execution times before the mapping (ETBM) deduced from [3] [15], the TDMA time slices allocated to each actor in the tile on which is mapped (TDMA) and the resulting execution times of actors after the mapping (ETAM) are summarized in Table 2.

Table 2: Actors mapping and TDMA allocations in the MJPEG decoder SDFG

| Tiles | Tile T1 | | Tile T2 | | Tile T3 | |
|---|---|---|---|---|---|---|
| **Actors** | **VLD** | **IZZ** | **IQ** | **IDCT** | **CC** | **RE** |
| **ETBM** | 2082463 | 24791 | 49582 | 99165 | 74374 | 892484 |
| **TDMA** | 50000 | 50000 | 10000 | 90000 | 20000 | 80000 |
| **ETAM (*)** | 2132463 | 74791 | 139582 | 109165 | 154374 | 912484 |

*Legend :*
ETBM = **E**xecution **T**ime **B**efore **M**apping (clk)
ETAM = **E**xecution **T**ime **A**fter **M**apping (clk)
(*) ETAM(actor) = ETBM(actor) + TDMA(other actors mapped on the same tile)

To model this proposed mapping, we used the binding-aware graph transformations already detailed above. The resulting SDFG of the mapped MJPEG decoder is presented in Fig. 10.

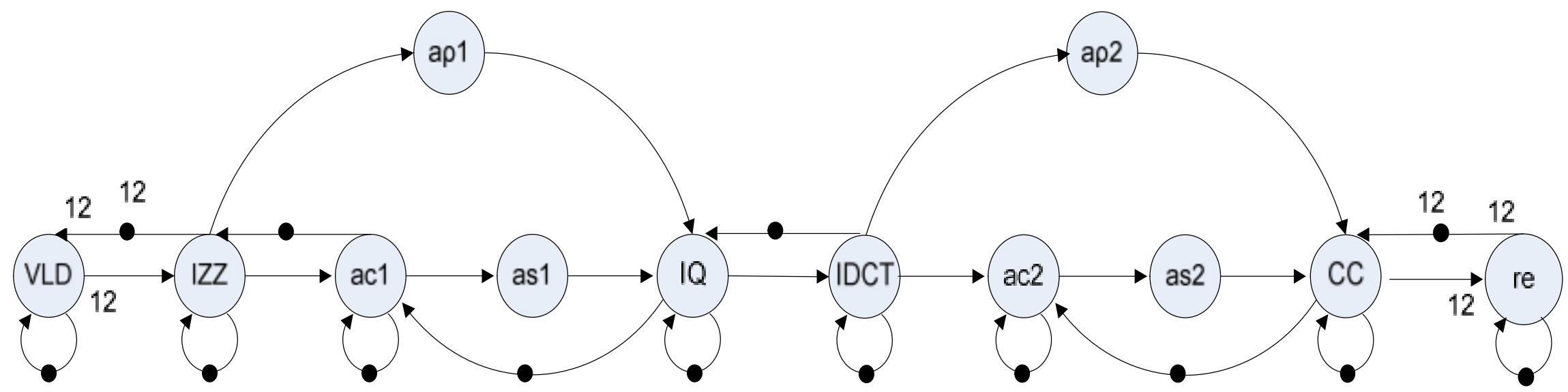

| Actor | VLD | IZZ | IQ | IDCT | CC | RE | $a_{c1}/a_{c2}$ | $a_{\rho1}/a_{\rho2}$ | $a_{s1}$ | $a_{s2}$ |
|---|---|---|---|---|---|---|---|---|---|---|
| **ET (clk)** | 2132463 | 74791 | 139582 | 109165 | 154374 | 912484 | 252047(*) | $10^{5}$(**) | 90000 | 80000 |

(*) $ET(a_{c1})$ = latency of the connection + [size of a communicated token/bandwidth of the connection] = L(c) + [sz/β]
= 3 + [1024/0.00406278] = 252047 time units.
$ET(a_{c2})$ = L(c) + [sz/β] = 3 + [512/0.00203139] = 252047 time units.
(**) $ET(a_{\rho1}) = TE(a_{\rho2})$ = latency calculated by the fifth step of the predictable design flow = ρ = 100000 time units.

Fig. 10 Mapped MJPEG decoder before migration

After presenting the SDFG of the mapped MJPEG decoder of Fig. 10 with the SDF3 XML format [16], we inject the resulting XML file to the throughput computation algorithm implemented by the SDF3 tool. The algorithm output is a throughput equal to 13,6 frames/second (f/s). We notice that obtained throughputs are computed for processors frequency of 100 MHz.

The most CPU greedy tasks of the MJPEG decoder are VLD (35%) and IDCT (20%) [3]. Therefore, we will respectively migrate the VLD and IDCT actors to hardware blocks. Then, we will evaluate the MJPEG decoder throughput after migration.

## 4.3 Mapped MJPEG decoder after VLD migration

Preserving the same mapping as before the migration, the SDGF of Fig. 11a models the MJPEG decoder having the VLD task migrated to a hardware block. The modeling of the migration impacts detailed above will be applied to this particular migration case:

• *Impact 1:* We assume that the execution time of the migrated VLD will be reduced to the half of its execution time before migration. Therefore, its execution time is henceforth equal to 1041231 = 2082463/2.

• *Impact 2:* Before the migration occurs, VLD and IZZ actors shared the processor of the tile T1. That is why IZZ firing is postponed by the TDMA time slice allocated to actor VLD. Once the VLD actor is migrated to a hardware block, the IZZ actor has no longer to wait the 50000 clock cycles reserved to VLD actor as TDMA time slice.

• *Impact 3:* Since actor IZZ requires a block of 8 by 8 pixels for its $(i+1)^{th}$ execution, this block can be entirely pre-fetched during its $(i)^{th}$ execution and the fetch mechanism is no more needed. So, the HS1 communication between VLD and IZZ actors is modeled by the memory-aware SDFG of Fig. 3b removing the m actor.

The throughput of the MJPEG decoder after the migration of the VLD actor is equal to 15.58 f/s. We remind that this throughput is obtained using images of 32*24 resolution and processors frequency of 100 MHz. The VLD hardware implementation allows the decoding of 2 extra frames per second compared to the decoding throughput before migration.

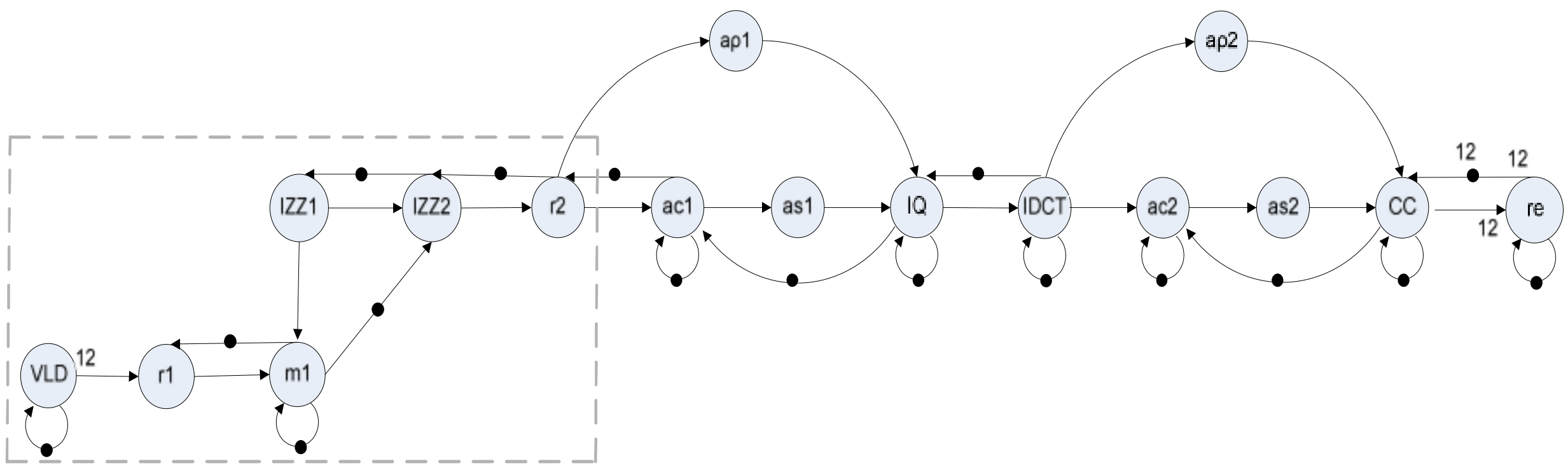

| Actor | VLD | IZZ2 | IZZ1 | IQ | IDCT | CC | RE | $a_{c1}/a_{c2}$ | $a_{p1}/a_{p2}$ | $a_{s1}$ | $a_{s2}$ | $r_1/r_2$ | $m_1$ |
|---|---|---|---|---|---|---|---|---|---|---|---|---|---|
| ET(clk) | 1041231 | 24791 | 10000 | 139582 | 109165 | 154374 | 912484 | 252047 | $10^5$ | 90000 | 80000 | 1 | 262047(*) |

(*) We suppose that the time necessary for the actor $m_1$ to pre-fetch data in the memory of the HW block is equal to 10000 time units. The latency required to transfer data pre-fetched from the HW block memory to the SW task that requests the data (IZZ2) is added to the execution time of the actor m1: ET(m1)=10000 + 252047.

(a) VLD Migration

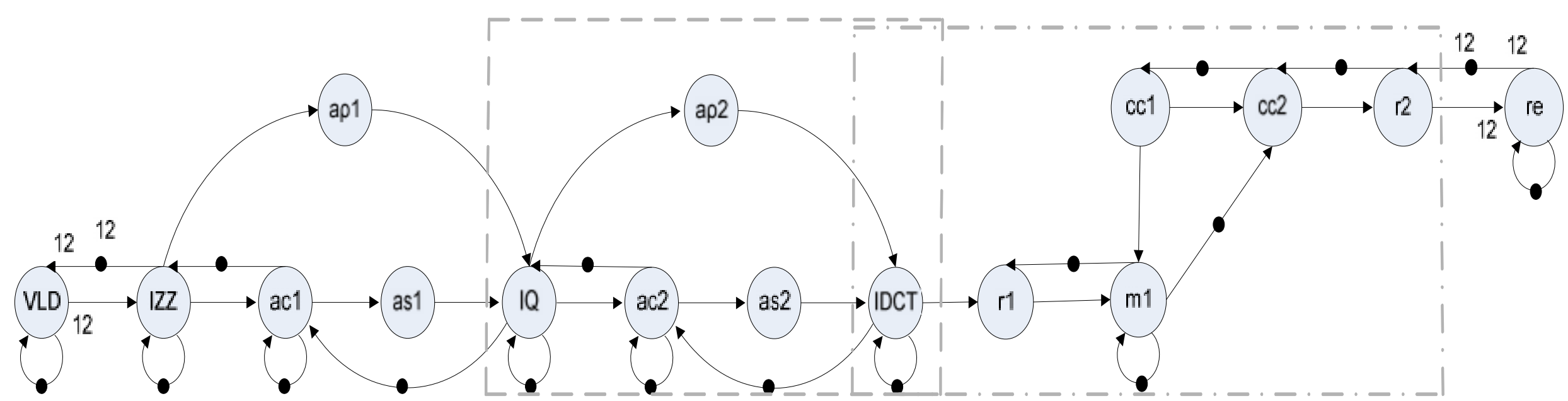

| Actor | VLD | IZZ | IQ | IDCT | RE | CC1 | CC2 | $a_{c1}/a_{c2}$ | $a_{p1}/a_{p2}$ | $a_{s1}$ | $a_{s2}$ | $r_1/r_2$ | $m_1$ |
|---|---|---|---|---|---|---|---|---|---|---|---|---|---|
| ET(clk) | 2132463 | 74791 | 49582 | 49582 | 912484 | $10^4$ | 154374 | 252047 | $10^5$ | 0 | 0 | 1 | 262047 |

(b) IDCT Migration

Fig. 11 Mapped MJPEG decoder after migration

## 4.4 Mapped MJPEG decoder after IDCT migration

The second considered case is the migration of the IDCT task to a hardware block. Fig. 11b presents the SDFG of the mapped MJPEG decoder after the IDCT migration as well as its actors execution times:

• *Impact 1:* The execution time of the IDCT actor after migration is equal to 49582 clock cycles (99165/2).

• *Impact 2:* The execution time of the IQ actor does not consider the 90000 TDMA time slice reserved to the IDCT actor since the latter is executed as a separate hardware block.

• *Impact 3:* The SH1 communication between the software actor IQ and the hardware actor IDCT is modeled by the binding-aware SDFG of Fig. 4b. The memory-aware SDFG is employed to model the HS1 communication between IDCT and CC actors.

The MJPEG decoder throughput after IDCT migration is equal to 17,23 f/s which mean a gain of about 4 f/s. Table 3 summarizes the obtained throughputs before and after migration.

Table 3: The Migration gain of the MJPEG decoder

| | | |
|---|---|---|
| Throughput without Migration (f/s) | 13,6 | |
| | **VLD** | **IDCT** |
| Throughput with Migration (f/s) | 15,58 | 17,23 |
| **Migration Gain (f/s)** | **1,98** | 3,63 |

## 5. Conclusions and perspectives

In this paper, we have proposed a SDF model that considers the software to hardware task migration impacts exploiting graphs used by the predictable design flow to model applications mapped to NoC-based MPSoCs. The proposed model was applied to the real multimedia application MJPEG decoder to estimate the migration performances in terms of throughput. The experimental results show that, using this model, the migration of VLD and IDCT tasks of the MJPEG decoder leads to respective throughput increases of about 2 and 4 frames per second.

The proposed migration model is particularly useful when designers want to take decisions for or against tasks hardware implementation during design phase. Once programmed, our proposition allows the migration performance estimation in clearly reduced deadlines (seconds) compared to its performance estimation using simulations (hours). This gain, at the level of design time, amounts to the high level migration modeling that we have proposed. Designers can explore several migration cases and estimate their performances in tiny delays to cope with the constantly increasing MPSoCs design complexity and time-to-market pressure.

We suggest as continuation to this work the implementation of the proposed migration model to automate its use and facilitate its applicability. Moreover, the migration model can be refined exploiting the Guaranteed Throughput Channel proposed in [11].

**Dorsaf SEBAI** received the Engineer degree from the National Institute of Applied Sciences and Technology, Tunisia in 2008 in networks and computer sciences. She received the Master degree from the Polytechnic School of Tunisia in 2009 in Electronic Systems and Communication Networks. Actually, she is working as Assistant at the National Institute of Applied Sciences and Technology in Tunis. She is preparing to begin her thesis.

**Abderrazak JEMAI** received the Engineer degree from the University of Tunis, Tunisia in 1988 and the DEA and “Doctor” degrees from the University of Grenoble, France, in 1989 and 1992, respectively, all in computer sciences. From 1989 to 1992 he prepared his thesis on simulation of RISC processors and parallel architectures. He became an Assistant Professor at the ENSI university in Tunis in 1993 and a Maitre-Assistant Professor at the INSAT university in Tunis since 1994. He was the principal investigator for the “Synthesis and Simulation of VLSI circuits” project at the ENSI/Microelectronic group, the simulation module in AMICAL at TIMA in Grenoble and the “Performance evaluation of MPSoC” project in LIP2/FST Laboratory in Tunis.

**Imed BENNOUR** received the Master and Ph.D. degrees in CS from the University of Montreal in 1992 and 1996. He has served as scientist at Nortel Telecom Company in Ottawa. He is currently working as Assistant Professor at the University of Monastir, Tunisia. His current research and teaching interests are mainly Systems on Chips design methodologies, hardware/software verification and high-level synthesis.